\documentstyle[epsfig,dina4,amssymb]{article}
\newcommand{\p}{\mbox{e$^+$}}
\newcommand{\n}{\mbox{e$^-$}}
\newcommand{\pn}{\mbox{e$^+$e$^-$}}

\newcommand{\UT}{\mbox{$^{238}$U + $^{181}$Ta\ }}
\newcommand{\UPA}{\mbox{$^{238}$U + $^{208}$Pb\ }}
\newcommand{\UPB}{\mbox{$^{238}$U + $^{206}$Pb\ }}
\newcommand{\UTH}{\mbox{$^{238}$U + $^{232}$Th\ }}
\newcommand{\UU}{\mbox{$^{238}$U + $^{238}$U\ }}
\newcommand{\Ta}{\mbox{$^{181}$Ta\ }}
\newcommand{\mug}{\mbox{\,$\mu g/cm^2$\ }}

\begin{document}
\begin{titlepage}
\hspace*{12.5 cm}{\bf UCY--PHY--96/13}
\vspace*{0.5cm}
\begin{center}
\LARGE
\bf
 \mbox{New results on \pn--line emission in U+Ta collisions}\\
\end{center}
 
\vspace*{0.5cm}
\begin{center}
\large
 {\bf The ORANGE Collaboration at GSI}

\vspace*{3.5 mm}
\Large
 U.~Leinberger$^a$, E.~Berdermann$^a$, F.~Heine$^b$, S.~Heinz$^b$, 
 O.~Joeres$^b$, P.~Kienle$^b$, I.~Koenig$^a$, W.~Koenig$^a$, 
 C.~Kozhuharov$^a$, M.~Rhein$^a$, A.~Schr\"oter$^a$, 
 H.~Tsertos$^{c,}$\footnote{Corresponding author, 
  e-mail "tsertos@alpha2.ns.ucy.ac.cy" \\
 Dept. Nat. Science, Univ. of Cyprus, PO 537, 1678 Nicosia, Cyprus}

\large
\vspace*{3.5 mm}
$^a$ Gesellschaft f\"ur Schwerionenforschung (GSI),
                                              D--64291 Darmstadt, Germany\\
$^b$ Technical University of Munich,  D--85748 Garching, Germany\\
$^c$ University of Cyprus, CY--1678 Nicosia, Cyprus\\ 
\end{center}
 
\vspace*{0.5cm}
\large
\begin{center}
\section*{Abstract}
\end{center}
\large        
We present new results obtained from a series of follow-up
e$^+$e$^-$--coincidence measurements in heavy-ion collisions, utilizing an 
improved experimental set-up at the
double-Orange $\beta$-spectrometer of GSI.
The collision system \UT was reinvestigated in three independent runs
at beam energies in the range (6.0$-$6.4)$\times$A\,MeV and different target 
thicknesses, with the
objective to reproduce a narrow sum-energy \pn--line at $\sim$635\,keV 
observed previously in this collision system.
At improved statistical accuracy, the line could not be found 
in these new data.  
For the "fission" scenario, an upper limit (1$\sigma$) on its production 
probability per collision of 
1.3$\times$10$^{-8}$ can be set which has to be compared to the 
previously reported 
value of [4.9$\pm0.8(stat.)\pm1.0(syst.)]\times$10$^{-7}$.
Based on the new results, a reanalysis of the old data 
shows that the 
continuous part of the spectrum at the line position is significantly
higher than previously assumed, thus reducing 
the production probability of the line by a factor of two and its 
statistical significance to $\le$3.4$\sigma$.
\end{titlepage}

\section{Introduction}

\large
Previous \pn -coincidence measurements in heavy-ion collisions at the
Coulomb barrier, performed at the UNILAC accelerator of GSI 
by the EPOS and ORANGE collaborations, have shown 
narrow \pn -sum-energy lines with energies in the range 550--810 keV~[1--4]. 
Their features were found to be complex, and did not fit into any
conventional atomic and/or nuclear production process.
In particular, the speculation that a hitherto unknown neutral 
particle (mass$\sim$1.8 MeV/c$^2$), decaying into e$^+$e$^-$ pairs, might be
involved~\cite{Cow86}, has conclusively been ruled out by subsequent 
Bhabha-scattering experiments~\cite{Tse91}.
Thus, the origin of this
phenomenon has remained a puzzle, without a satisfactory explanation until now.

\vspace*{2mm}
\hspace*{3mm}
On the other hand, all the data reported previously were incomplete as far as
a systematic dependence on the collision parameters and the lepton emission
scenario is concerned, and were suffering from limited statistical significance.
Even more, a consistent description of the line characteristics (i.e., energies,
cross sections) could not be achieved by comparing the results of both groups.
For instance, the cross sections of the lines reported by us for \UU\ and \UPA\
collisions \cite{Koe89,Ikoe93} were found to be an order of magnitude
smaller than those quoted by the EPOS collaboration \cite{Cow86,Sal90} and, in
particular, a line at $\sim$760\,keV, initially reported by EPOS 
for \UTH\ collisions \cite{Cow86}, has not been observed in our experiments.
From our investigations, the most distinct evidence exists for a sum-energy
line at $\sim$635 keV observed in the collision system \UT by using
a beam energy of 6.3$\times$A\,MeV and 1000 \mug thick \Ta target \cite{Ikoe93}.
The line was seen with the so far
highest statistical significance (6.5$\sigma$)
by selecting \pn -pairs in coincidence with
two heavy ions (HI), whose kinematics is consistent with fission
of the U after the collision. 
However, the opening-angle distribution of the \pn--pairs associated 
with this line, as measured
directly in our experiments, was found to be rather isotropic, whereas 
the line energy is unequally shared
between positrons and electrons, thus being in clear disagreement with
a two-body decay scenario suggested previously \cite{Cow86,Koe89}. 

\vspace*{2mm}
\hspace*{3mm}
It is obvious that further experiments, which could be able to
clarify this rather unsatisfactory situation, were urgently needed.
We started a new round of experiments
by upgrading the detection systems at the double-Orange spectrometer
\cite{Bae95} 
and exploiting the new high-charge state injector at the UNILAC accelerator.
Here we report the first results from the new investigations, which explore
with significantly improved sensitivity, the collision system \UT.

\vspace*{5mm}
\section{Experimental set-up}

\large
As in previous experiments \cite{Koe89,Ikoe93}, we used
two identical iron-free, orange-type $\beta$- spectrometers,
facing each other with a common object point, at which a rotating
target wheel is placed (Fig. 1). 
Positrons (\p) emitted in the backward 
($\theta_{e^+} = 110^{\circ} - 145^{\circ}$) 
and electrons (\n) emitted in the forward
hemisphere ($\theta_{e^-} = 38^{\circ} - 70^{\circ}$)
are focussed onto position sensitive detectors.
Each lepton  detector consists of an array of high-resolution 
Si PIN diodes, called "PAGODA".
Intrinsic to this set-up is the capability of focussing, at  
a given field setting, only a certain momentum interval of \p{} or \n{}
by rejecting at once the opposite charge.
This is a major advantage because the high $\delta$-electron
background is therewith suppressed completely on the \p--side, while
the sharp  low-energy cut-off
on the \n--side enables operation at high luminosities. 

\vspace*{2mm}
\hspace*{3mm}
The leptons are identified by matching their momentum, derived
from the hit-point on the PAGODA and the spectrometer field setting,
with their energy, measured by the PIN diodes.
This provides a clear signature for \p{} and \n{}, and 
suppresses backgrounds due to $\gamma$ rays and scattered $\delta$-electrons
efficiently \cite{Lein96}.
An additional coincidence requirement with the 511~keV annihilation radiation
is thus not necessary for the \p{} identification.
The momentum-energy matching in addition rejects leptons backscattered from 
the PIN diodes almost completely, a unique feature of this set-up.

\vspace*{2mm}
\hspace*{3mm}
The old lepton detection systems \cite{Koe89,Ikoe93} are replaced by new
ones each consisting of 72 segmented, high-resolution, 1 mm thick, Si PIN 
diodes which are mounted on 12 roofs of a pagoda-like structure,
6 detectors on each roof (see Fig. 1).
Each detector chip has a trapezoidal shape
(24 mm base, 16 mm top, and 16 mm height) and is
tilted by $\sim 40^\circ$ relative to the spectrometer axis.
Each detector is further subdivided into three electrically separated 
sectors, covering an azimuthal angular range of $\Delta\phi_{lept}$ = $20^\circ$.
A set of three neighbouring sectors at the same azimuthal angle is
connected to one preamplifier, recording information on energy and arrival
time of a lepton.
To identify the individual sector of the set, the rear of each
sides of all detectors on each roof are read out additionally by another
preamplifier.
The e$^+$e$^-$-sum-energy 
and time resolution achieved in-beam amounts to $\sim$15 keV
and $\sim$4 ns (FWHM), respectively, cooling down the detectors to
$\sim-$25$^\circ$C. This is comparable to
that of our old set-up.
       
\vspace*{2mm}
\hspace*{3mm}
The opening angle of the \pn--pair, $\theta_{e^+e^-}$, is measured directly
within a range of
$40^{\circ}-180^{\circ}$ in the laboratory. Using the
$\phi$--separation of the PAGODA's, this range can be subdivided into 
ten angular bins with a typical width of  $\sim\pm 10^{\circ}$ and
centroids at 
$\theta_{e^+e^-}$~=~70$^{\circ}$, 73$^{\circ}$, 80$^{\circ}$, 90$^{\circ}$,
102$^{\circ}$, 116$^{\circ}$, 131$^{\circ}$, 145$^{\circ}$,
159$^{\circ}$, and 167$^{\circ}$. Thus, the
lepton opening angles are measured with an accuracy which is comparable to the
effect of small-angle scattering in the targets
(e.g. $\sim$24$^\circ$ for a 1\,mg/cm$^2$ thick target \cite{Ikoe93}).
Each pagoda array covers a maximum momentum acceptance of
$\Delta p/p$ =30\% which corresponds to an energy interval 
of $\Delta E \sim$150 keV  at a lepton energy of  300 keV.
Within this momentum interval, the full-energy peak efficiency 
is 10\% and 11\% of 4$\pi$, for electrons and positrons, respectively. 
With respect to the e$^+$e$^-$-coincidence efficiency and the range of
energy-difference distributions covered, this is an improvement 
by about a factor of two as compared to our previous set-up.

\vspace*{2mm}
\hspace*{3mm}
In addition, a modified set of heavy-ion counters (PPAC's)
measures the polar angles
of the scattered ions in the ranges
$13^{\circ} \leq  \theta _{ion} \leq 35^{\circ}$ and
$40^{\circ} \leq  \theta _{ion} \leq 70^{\circ}$ 
with an accuracy of 1$^{\circ}$ and 0.5$^{\circ}$, respectively.
The total azimuthal range is covered with a resolution in 
$\phi_{ion}$  of $20^\circ$.
Finally, a high-resolution Ge(i) detector is used to measure $\gamma$ rays
(E$_{\gamma} \gtrsim$ 1 MeV) in coincidence with the scattered heavy ions 
(see Ref. \cite{Ikoe93}).

\vspace*{2mm}
\hspace{3mm}
The performance of the lepton detection systems has been studied extensively
in source measurements \cite{Lein96}.
In particular, the \pn--opening-angle resolving power is demonstrated by
measuring the 1.76 MeV E0 transition in $^{90}$Zr \cite{Bae95,Lein96} and the
1.77 MeV M1 transition in $^{207}$Pb \cite{Lein96}.
Even more important is the check of the capability of the set-up 
to detect narrow sum-energy lines under beam conditions. 
This has been proven by measuring the 1.844 MeV E1 transition in $^{206}$Pb,
populated via Coulomb excitation in \UPB collisions. 
By applying an event-by-event Doppler-shift correction assuming emission from
the recoiling $^{206}$Pb
nucleus, a weak \pn--sum-energy line at $\sim$820\,keV is observed, whose 
intensity is consistent with the yield expected from the corresponding 
$\gamma$ line at 1.844 MeV measured with the Ge(i) detector \cite{Sophiepriv}.

\vspace*{2mm}
\hspace{3mm}
A considerable improvement in these experiments is achieved by exploiting
the new high-charge state injector with an ECR ion source at the
UNILAC accelerator. 
The latter provides beams with an improved time structure 
(9 ns between micropulses compared to 37 ns of the old injector), which
allows to double luminosity
at a random coincidence rate reduced by a factor of two due to the lower 
micropulse intensity. Random coincidences were further reduced by the almost
complete lack of plasma oscillations in the new source.

\vspace*{5mm}
\section{Experiments and results}

\large
With the new set-up we reinvestigated the system \UT\ in three independent runs,
using beam energies of 6.0, 6.1, 6.15, 6.3, and 6.4$\times$A\,MeV and target
thicknesses of 600, 800, 1000, and 1200\mug. These runs, summarized in 
table \ref{expt_conds}, completely cover the kinematical parameter 
space of the previous experiment.
In the first of the new experiments the old injector was used, providing a 37\,ns 
microstructure
of the beam as it was the case in the old experiments \cite{Ikoe93}. 
The later runs were carried out with beams from the
new injector with an improved  microstructure of 9\,ns.

\vspace*{2mm}
\hspace{3mm}
The main goal of much better statistics could be reached successfully,
as it is shown in table \ref{expt_conds}.
In the following we concentrate on experiments
with incident beam energy of 6.3$\times$A\,MeV, since the line to be reproduced
was found at this beam energy. 
A comparison of the field settings in the various dipole magnets in
the beam line, used as a cross-check of the accelerator's independent beam
energy measurement, exhibited no significant differences ($<$1\%) between the 
old and new experiments. 
Also the acceptance for both the heavy ions and the leptons
has been essentially the same in all experiments discussed here.

\vspace*{2mm}
\hspace{3mm}
We analyzed the new data by applying the same  conditions as in
the old experiment; i.e., all \pn\ opening angles, a cut in the  
\pn-energy differences between  $-$66 keV and 150\,keV, and two
coincident heavy ions with either peripheral quasi-elastic kinematics
($18.6\,fm < R_{min} < 29.3\,fm$), or 
non-two-body kinematics  covering the kinematical range of fission
in two opposite 60$^\circ$ $\phi$-segments of the HI detector with 
$15^\circ <\theta_1 < 25^\circ$ and $40^\circ <\theta_2 < 70^\circ$.
None of the resulting \pn-sum-energy spectra revealed a line
with the production probability per collision as previously deduced from the
old experiment \cite{Ikoe93}.
As an example, we show in Fig. 2   
the corresponding sum-energy spectra, obtained from non-two-body heavy-ion 
collisions (''fission scenario''),  superimposed with the expected line
intensities from the old result \cite{Ikoe93}.

\vspace*{2mm}
\hspace{3mm}
The 1$\sigma$ upper limits for the production probability
of a line at $\sim$635\,keV, derived from the new
results, are much lower than the reported production probability from the 
old experiment \cite{Ikoe93}. 
These results are summarized in table \ref{expt_lims}. 
As can be seen, a summation of all the new results, taken at the beam energy 
of 6.3$\times$A\,MeV, yields upper limits  which are a factor of 
15 (quasi-elastic collisions) and 37 (non-two-body collisions)
smaller than the production probabilities derived from the old experiment.

\vspace*{2mm}
\hspace{3mm}
Detailed studies of the new experiments showed that the shape of the 
continuous part of the sum-energy
spectra for the fission scenario can be well described 
by the corresponding sum-energy spectra, measured in coincidence 
with quasi-elastic
collisions, and smooth distributions generated with an event-mixing
technique from the latter. This was a surprising new result, since different
reaction kinematics can well produce different continua in the sum-energy
spectra.

\vspace*{2mm}
\hspace{3mm}
The various experiments were found to be generally in good agreement.
In particular, we were able to understand and quantify a class of
scattering events in the inner HI detector, which obscures 
the signature of non-two-body HI kinematics consistent with fission.
The number of peripheral quasi-elastic collisions 
normalized to non-two-body collisions
(fission) was found to be $\sim$3 and $\sim$9, for
the data taken with the old and new injector, respectively.
Detailed investigations have revealed that this is caused by a 
strong contamination of misidentified double hits in the inner PPAC,
where three heavy ions from the same micropulse give a misleading signature of 
the fission scenario. 
In this case, a quasi-elastically scattered ion pair in the inner and
outer counter, as well as an independently scattered ion in the inner counter,
are detected simultaneously. 
Two of these ions hit the inner PPAC, thus
causing loss of the unambiguity of the ($\theta,\phi$)--matrix identification, 
within it's time resolution.
A reanalysis of the old data showed that nearly $2/3$ of the
events in the sum-energy spectra (including also events of the peak),
have to be attributed to such misidentified  ''fission events''.
In the data taken with the new injector, this contamination is reduced by
more than a factor of three \cite{Lein96}.

\vspace*{5mm}
\section{Discussion of the results}

\large
From our new investigations of the collision system \UT, carried
out with improved experimental sensitivity, we have not found 
evidence for the previously reported line \cite{Ikoe93}
in any of the new data.  
This negative outcome represents therefore a serious discrepancy to our 
first measurement in this collision system, being obtained  
with lower statistical accuracy \cite{Ikoe93}.

\vspace*{2mm}
\hspace{3mm}
Since the relevant kinematical parameter space has largely been covered and
extended by the new experiments (see table 1), and since possible target
deterioration effects have been kept lower, or at most comparable to the
previous measurements, we cannot attribute the absence of the line in the new
experiments to an incorrectly chosen beam energy or deteriorated targets.
Furthermore, detailed studies of the in-beam performance of our new set-up
have not provided any evidence for a possible incapability of detecting narrow
sum-energy lines.
The signal/background ratio
in the measured \p and \n spectra was found to be comparable to the old set-up.
The production probabilities of the measured continuous spectra, around 
the expected line position,  were found to be consistent in all the 
new measurements, and their values agree within 18\% with
those obtained from the old experiment \cite{Lein96}.
This is also the case for the number of fission events normalized
to the number of scattered heavy ions, when misidentified double hits 
(discussed above) are accounted for.
The detection efficiency for fission fragments has been similar in all
experiments. 

\vspace*{2mm}
\hspace{3mm}
From the discussion above, we have to conclude that it is 
difficult to find a physics-based working hypothesis which would corroborate
both, the appearance of the line in the previous results and its absence
in the new, improved experiments.
Taking into account that we have not found any evidence that the reported line
might be due to trivial effects or background processes, its statistical
significance has to be reconsidered. 

\vspace*{2mm}
\hspace{3mm}
In Fig. 3 we show the \pn\ sum-energy spectrum with the reported line 
at $\sim$635 keV
\cite{Ikoe93}, together with a polynomial fit (solid line) used at that time
to describe the continuous part of the spectrum outside the line.
Based on this background curve, the statistical significance of the line was
deduced to  $\sim$6.5$\sigma$ \cite{Ikoe93}.
However, recent reanalysis of these results has revealed that the formerly
used polynomial fit might be incorrect, as demonstrated by the new 
dashed background curve. The latter has been gained by
making use of the new observation that the shape of the continuous part of the
spectrum of non-two-body events (fission scenario) can be  well described by
event mixing of \p\ and \n\ obtained from quasi-elastic scattering
(see sect. 3).
Taking this background curve as a reference, the remaining excess around 
635 keV has a statistical significance of $\lesssim$3.4$\sigma$, and  
can be translated into a production probability of
[2.7$\pm 0.8(stat.)\pm 0.6(syst.)]\times 10^{-7}$.
The corresponding value for peripheral quasi-elastic collisions
amounts to [2.7$\pm1.2(stat.)\pm 0.6(syst.)]\times10^{-7}$. These values are
approx. by a factor of two lower than those reported in the 
previous analysis (see table 2).
 
\vspace*{2mm}
\hspace{3mm}
We shall note in this context that in some runs we found an indication 
for a narrow \pn- line in the sum-energy spectra by applying slightly 
different cuts in the HI kinematics (see Ref. \cite{Lein96}).
This line, however, could not be reproduced
in the subsequent runs taken under similar experimental conditions and
better statistics. 
Adding all data together taken at 6.3$\times$A MeV, we  
derive upper limits for the production probabilities
of narrow lines at $\sim$635\,keV of $P(1\sigma)= 2.9\times 10^{-8}$ for
quasi-elastic collisions and 
$P(1\sigma)= 1.3\times 10^{-8}$ for the fission scenario, normalized to
peripheral quasi-elastic collisions.
Finally, it should be pointed out that negative results 
were also reported by the APEX \cite{APEX1} and EPOS \cite{Gan96}
collaborations, who focused on previous findings of
old EPOS measurements \cite{Cow86,Sal90}.  

\vspace*{1.5cm}
\large
{\bf Acknowledgement:}
{\em We would like to thank all the people of the UNILAC accelerator
 operating crew  for their efforts in delivering stable $^{238}$U
 beams with high intensities.}
 
\newpage
\large

\newpage
\large
\begin{center}
\subsection*{\bf TABLE CAPTIONS}
\end{center}
 
\vspace*{1.0cm}
\begin{table}[ht]
\caption{\large Summary of our new investigations in the collision system 
\UT (I--III).
For comparison, the old data from the same collision system are also
shown. 
As can be seen, the new data are obtained with significantly improved
statistics}
\label{expt_conds}

\vspace{5mm}
\large
\begin{center}
\begin{tabular}{|c|c|c|c|c|}\hline
           &            &             &            &                \\
\UT        &Beam energy &Targ. thick. &Beam struct.& True \\
           & [MeV/u]     & [$\mu$g/cm$^2$]& [ns] & e$^+$e$^-$--pairs\\\hline\hline
Expt. I & 6.3  & 1200 & 37 & 22900 \\
	& 6.3  &  600  & 37 & 14000 \\\hline
Sum	& \multicolumn{3}{|c|}{}			& 36900	\\\hline\hline

Expt. II& 6.3  &  800   & 9 & 43200 \\ 
	& 6.4  &  800   & 9 & 16900 \\
	& 6.1  &  800   & 9 & 9900 \\\hline
Sum	& \multicolumn{3}{|c|}{}		       & 70000	\\\hline\hline
Expt. III& 6.3 & 1000  & 9 & 57000 \\
	& 6.15 & 1000  & 9 & 51000 \\
	& 6.0  &  600  & 9 & 40000 \\\hline
Sum	& \multicolumn{3}{|c|}{}			&148000	\\\hline\hline
Total I--III & \multicolumn{3}{|c|}{}		&254900 \\\hline\hline
Old expt. & 6.3  & 1000 & 37 & 11200 \\\hline
\end{tabular}
\end{center}
\end{table}

\newpage
\large
\vspace*{1.0cm}
\begin{table}[ht]
\caption[bla]{\large 1$\sigma$ upper limits for the production probability per
collision, P$_{max}$(1$\sigma$), for a narrow sum-energy \pn--line 
at $\sim$635 keV in \UT collisions at a beam energy of 6.3$\times$A\,MeV.
Also shown are the line production probabilities derived from the old 
experiment \protect\cite{Ikoe93}.
}
\label{expt_lims}

\large
\begin{center}
\begin{tabular}{|*{7}{c|}}\hline
               &        &       &         &          &    & \\
\UT & Expt. I&Expt. I& Expt. II & Expt. III &  Old expt. & All Data \\
 E$_{beam}=$ 6.3$\times$A\,MeV &      &     &     &          &    & \\\hline
Targ. thick. [\mug] & 1200 & 600 & 800 & 1000 & 1000& \\\hline\hline
\multicolumn{7}{|c|}{Peripheral quasi-elastic collisions}\\\hline
P$_{max}$(1$\sigma$) [10$^{-8}$] & 7.4 & 8.6 & 4.5 & 4.5
&47$\pm12^a\pm10^{b\dagger}$&2.9\\\hline\hline
\multicolumn{7}{|c|}{''Fission'' normalized to peripheral quasi-elastic
collisions}\\\hline
P$_{max}$(1$\sigma$) [10$^{-8}$] & 4.5& 5.2 & 1.5 & 1.5 &49$\pm8^a\pm
10^{b\dagger}$&1.3\\\hline
\end{tabular}
\end{center}
\vspace{-2 mm}
\normalsize 
$^a$ statistical and $^b$ systematical uncertainty\\
\normalsize $^\dagger$ A recent analysis of these results
yielded production probabilities for the line
structure in case
of peripheral quasi-elastic collisions of 
[2.7$\pm 1.2(stat.)\pm 0.6(syst.)]\times10^{-7}$ and 
[2.7$\pm 0.8(stat.)\pm 0.6(syst.)]\times 10^{-7}$ for the fission
scenario.
\end{table}
%
%
\newpage
\large
\begin{center}
\subsection*{\bf FIGURE CAPTIONS}
\end{center}
 
\vspace*{1.0cm}
{\bf Fig. 1:}
Schematic view of the new experimental set-up at the double-Orange
device. 
Each of the $\beta$-spectrometers is
equipped with a $\beta$-multidetector system of 72 Si (PIN) diodes
(e$^{\pm}$--Pagodas).
The forward spectrometer is
surrounded by 18 position-sensitive heavy-ion detectors (PPAC), and
contains a further PPAC detector in its center.
Also shown is the rotating target wheel and the Ge(i) $\gamma$-ray detector.

\vspace*{1.0cm}
{\bf Fig. 2:}
Summary of e$^+$e$^-$-sum-energy spectra obtained from inelastic
U+Ta collisions at 6.3$\times$A\,MeV, leading to fission of the U nucleus, by
different target thicknesses indicated. 
The data shown in {\bf a)} and {\bf b)} are from the run I, in {\bf c)} 
and {\bf d)} from the runs II and III, respectively (see table 1).
The spectra are integrated over the whole \pn--opening-angle range
covered experimentally. 
The selection criteria (see text) are the same as those
imposed on the previous results \cite{Ikoe93}.
The superimposed peaks (dashed lines) at $\sim$635 keV would correspond 
to the signal 
expected from the line intensity reported previously \cite{Ikoe93}.
The difference in the signal/background ratios shown is due to the 
different contamination in misidentified double hits, discussed in the text.
                                                               
\vspace*{1.0cm}
{\bf Fig. 3:}
e$^+$e$^-$-sum-energy spectrum from U+Ta collisions at 6.3$\times$A\,MeV, 
obtained previously \cite{Ikoe93}. In this spectrum a prominent \pn--line at
$\sim$635 keV is seen.
The solid curve is a simple polynomial fit to the continuous part of 
the spectrum used in the first analysis \cite{Ikoe93}.
The dotted-line distribution is based on recent reanalysis of these 
data by using event mixing (see text).

%
%
\newpage
\pagestyle{empty}  

\vspace*{3cm}
\begin{center}
\epsfig{file=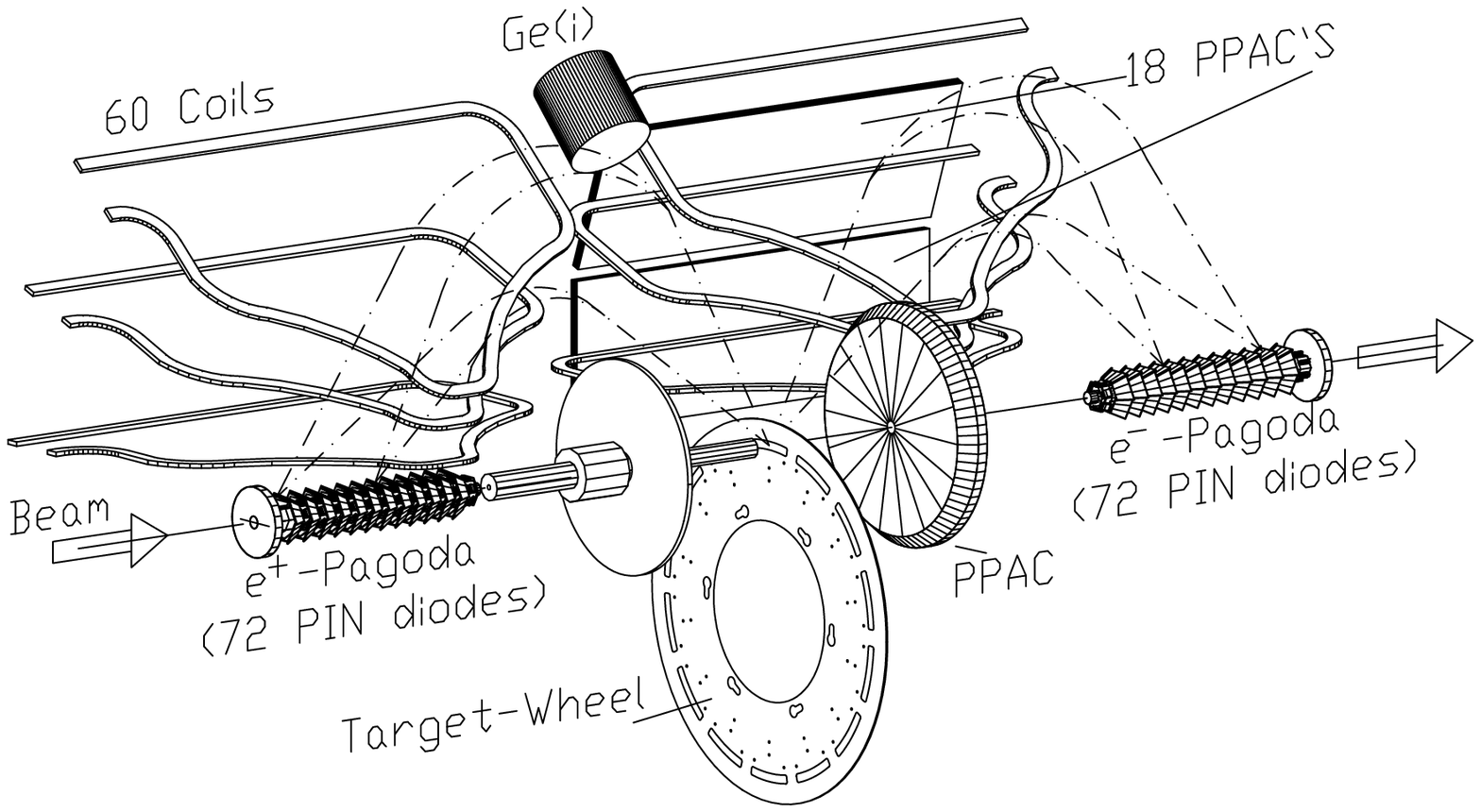,width=1.1\linewidth}
\end{center}

 \vspace*{6cm}
 {\Large \bf Figure 1} 

\newpage

\vspace*{1cm}
\begin{center}
\epsfig{file=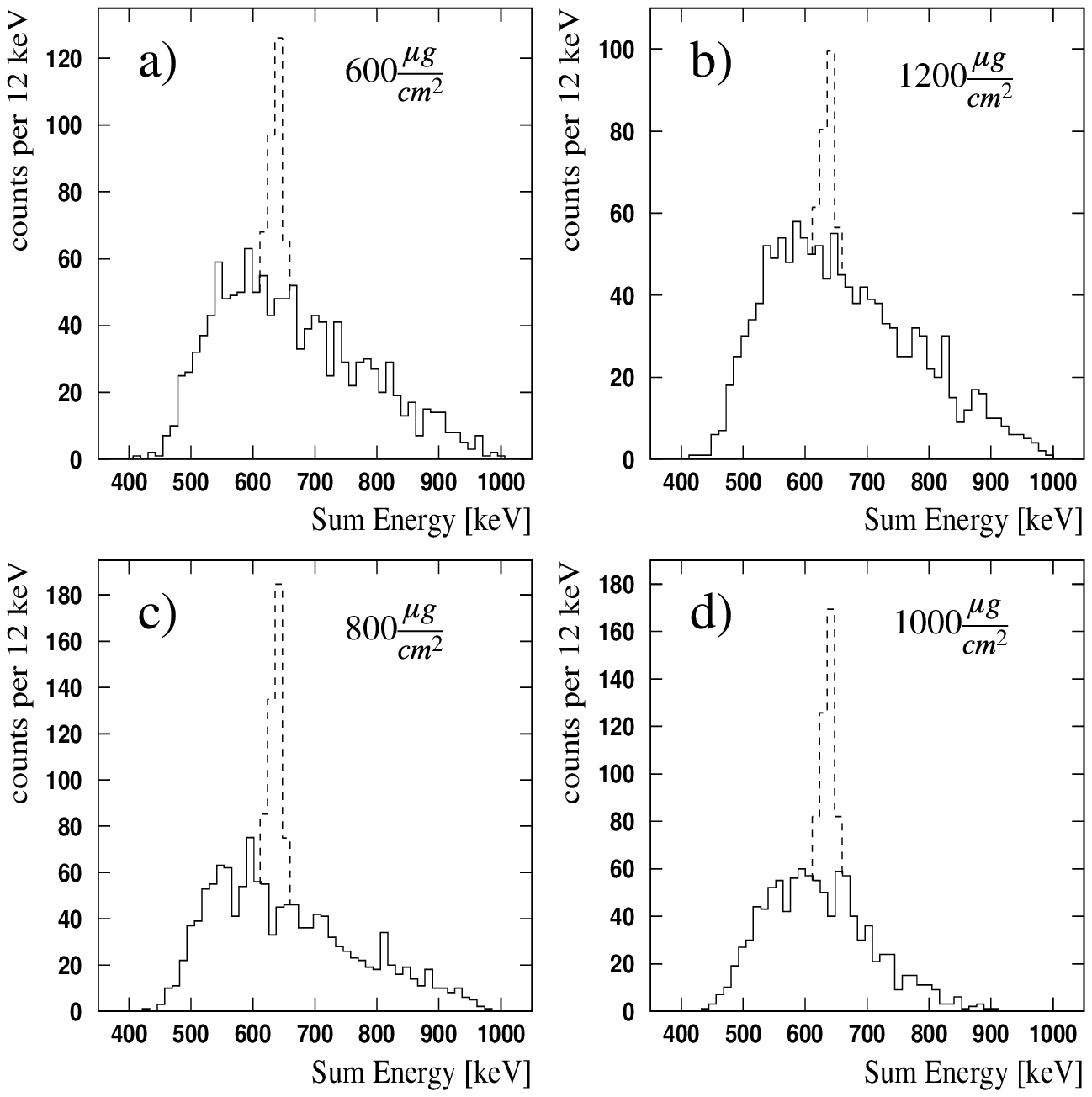,width=1.0\linewidth}
\end{center}

 \vspace*{3cm}
 {\Large \bf Figure 2} 

\newpage

\vspace*{1cm}
\begin{center}
\epsfig{file=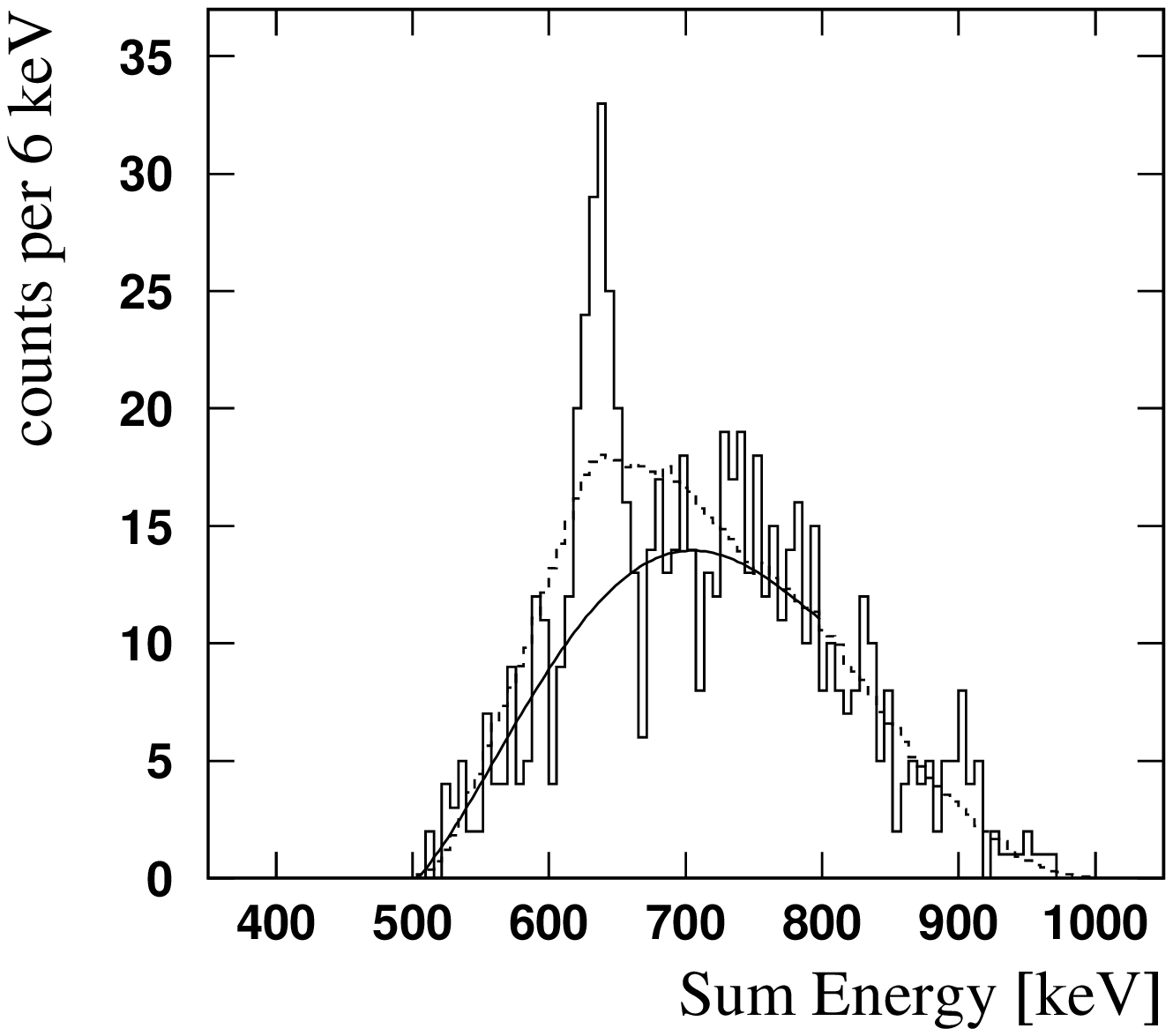,width=1.0\linewidth}
\end{center}

 \vspace*{3cm}
 {\Large \bf Figure 3} 

\end{document}